\documentstyle[prb,aps]{revtex}

\begin{document}

\title{Two-subband system in quantizing magnetic field: probing
 many-body gap by non-equilibrium phonons}
\author{ V.M. Apalkov and M.E. Portnoi\cite{byline}}
\address{
School of Physics, University of Exeter, Stocker Road, Exeter EX4 4QL, 
United Kingdom}

\maketitle
%\voffset=2.0true cm

\begin{abstract}

 We study the many-body effects in a two-subband quasi-two-dimensional 
 electron system in a quantizing magnetic field at filling factor
 three. A manifestation of these effects in the phonon absorption 
 spectroscopy is discussed.  
 The electron system is mapped onto a  two-level 
 system with the separation between levels determined by the intersubband 
 splitting and the cyclotron energy. The electron-electron interaction 
 enhances the excitation gap, which exists at all values of 
 the interlevel splitting. This results in a single-peak structure of 
 the phonon absorption rate as a function of magnetic field, instead of 
 the double-peak structure for non-interacting electrons.

\end{abstract}


\section{Introduction}

 The behavior of a two-dimensional (2D) electron system in a 
 quantizing magnetic field is completely determined by the interaction 
 part of the Hamiltonian. This results in  several interesting 
 phenomena in such a system, for example, including the Fractional 
 and Integer Quantum Hall Effects.\cite{qhe}
 A wide range of different experimental techniques has 
 been used to study this  system and its excitations.
 One of these methods is the  acoustic-phonon 
spectroscopy.\cite{phonon1,phonon2,phonon3,phonon4} 
 The specific feature of the electron-phonon interaction in 
 the 2D system in a magnetic field is a strong suppression of phonon induced 
 transitions with the phonon momentum larger than the 
 inverse magnetic length, $q_0 \approx 1/l$.\cite{levinson} 
 Due to the energy conservation the single-phonon transitions are only 
 possible if  $E(q_0) \approx \hbar s q_0$, where
 $s$ is the speed of sound and $E(q)$ is the dispersion of the 
 neutral excitations of the electron system. By changing $E(q_0)$ one
 can switch off or on acoustic phonon effects in the 
 magnetically-quantized 2D system. 

 At certain conditions a sharp magnetic field dependence of the 
 electron-phonon interaction can be achieved in a two-subband system.
 When the energy 
 splitting between the first and the second subbands 
 becomes close to the cyclotron energy the second Landau 
 level of the first subband has almost the same energy as the first 
 Landau level of the second subband. An interesting feature of this
 system is that the separation between these close Landau levels, $\Delta $, 
 can be changed by changing the magnetic field. Under these conditions 
 if there are electrons in the second Landau level of the first
 subband, i.e. the electron filling factor is greater then two, the
 phonon-induced effects become strong when $\Delta \approx \hbar s/l$. 
 This results in a double-peak structure of the dissipative electron 
 conductivity as a function of $\Delta $ or as a function of magnetic 
 field.\cite{misha1} The increase of the electron density in this system 
 increases the role of the  electron-electron interaction which  
 results in the renormalization of the excitation gap $\Delta $ and has 
 a considerable effect on the electron-phonon coupling. 
 In what follows we  analyze the case when the total 
 electron filling factor is equal to 
 three. We show that in this case the excitation spectrum has a
 gap at all values of interlevel splitting  $\Delta $. As a result 
 the acoustic-phonon absorption of the two-subband system has a single peak 
  as a function of the magnetic field. By lowering the electron 
 density one should observe the transformation of a single-peak 
 dependence into a double-peak one, which corresponds to a non-interacting
 case.

 We consider below the quasi-2D electron system confined in a single  
 heterojunction. 
 The effects related to the intersection of different Landau levels 
 belonging to different subbands were studied both theoretically 
 and experimentally for double quantum well structures.\cite{dw_e} 
 Unlike the double-well case the intersection of levels in the  
 two-subband single heterojunction  occurs at much higher magnetic
 field, which makes all field-dependent many-body effects more pronounced.

\section{Ground State and Excitations}

 At filling factor $\nu _0 =3$ the electrons completely occupy the 
 lowest Landau level (with both spin directions) of the first 
 subband. The only effect of the electrons in this  level on the electrons 
in the higher levels is the exchange renormalization of the interlevel 
splitting  $\Delta $. Because 
 this renormalization does not depend on the momentum of the electron 
 in the next levels in what follows we 
do not take the background (completely occupied) level into account  
  and consider $\Delta $ as the value,  which already includes 
the renormalization 
 due to the background electrons. As a result the system becomes the 
 two-level electron system with the effective filling factor $\nu =1$ and 
 the splitting between levels $\Delta $, see Fig.1. These levels are 
 the second Landau level ($n=1$) of the first subband and 
 the first Landau level ($n=0$) of the second subband. Both levels have 
 the same spin, e.g., $S_z = +1/2$ for a negative $g$-factor.

  It should be mentioned that the mapping of the $\nu _0 = 3$ system
 onto the two-level problem is valid for 
 any value of Zeeman splitting.  Due to exchange 
 interaction the ground state is a mixture of the 
 states with the same spin only.  Since we are interested in 
 the phonon induced transitions, which conserve 
 the electron spin,  we should consider the excitations without 
 spin reversal. As a result only the electron states with the same spin are
 relevant to the considered problem. 
   Because both levels have the same spin  
 we can treat electrons as spinless particles. For 
 convenience the levels are numbered by the 
 corresponding subband numbers, $\mu =1$ or $2$. To simplify the expressions
 the energy is measured in the Coulomb units $\varepsilon _C = e^2/\kappa l$ 
 and the length is in units of magnetic length $l$.
 Following the standard procedure\cite{kallin} 
 of rewriting the interaction part of the 
  Hamiltonian in the momentum representation, the total Hamiltonian of 
 the two-level electron system can be written in the form: 
\begin{eqnarray}
H & & =   \frac{\Delta }{2} \sum_{\mu \sigma k_y} (2\mu -3)
     C^{+} _{k_y \mu \sigma}
 C _{k_y \mu \sigma} +                    \nonumber  \\
& &      
 + \frac{1}{2}   \sum_{ \left\{\mu \right\} } 
  \sum_{q_x, q_y} \tilde{V}^{(n_1 n_4 n_3 n_2)}_{(\mu _1 \mu _4 \mu _3 \mu _2)} (\hat{q}) 
   \sum _{k_1,k_2} e^{i q_x (k_1-k_2)} C^{+} _{k_1+q_y, \mu_1} 
     C^{+} _{k_2, \mu_2}  C _{k_2+q_y, \mu_3} C _{k_1, \mu_4 }   \mbox{\hspace{3mm},}
\end{eqnarray}
  where the Landau level number is equal to $n_i = 1$ for $\mu _i =1$ and 
 $n_i = 0$ for $\mu _i = 2$, i.e. $n_i = 2-\mu_i$. 
We use the Landau  gauge with vector potential $\vec{A} = (0,Bx,0)$ and 
 $ C^{+} _{k, \mu }$, $ C _{k, \mu } $ are the creation and
 annihilation operators of the electron in the state $\psi _{n,k_y, \mu }$:
\begin{equation}
\psi _{n,k_y, \mu } (x,y,z) = \chi _{\mu} (z) \frac{e^{i k_y y}}
{\sqrt{L_y}} \phi _n (x-k_y)    \mbox{\hspace{3mm},}
\end{equation}
 where $\chi _{\mu } (z)$ is the envelope function of the $\mu$th subband;
 $\phi _n (x)$ is the $n$th harmonic oscillator function. 

 In Eq. (1) we  introduced the notations:
\begin{equation}
\tilde{V}^{(n_1 n_4 n_3 n_2)}_{(\mu_1 \mu_4 \mu_3 \mu_2)}(\hat{q}) ~=~ 
\frac{1}{q}~
         F_{\mu_1 \mu_4 \mu_3 \mu_2}(q) ~G _{n_1 n_4}(\hat{q}) 
     ~G _{n_3 n_2}(\hat{q}^*)   
      ~\exp\left ( -\frac{q^2}{2} \right)   \mbox{\hspace{3mm},}
\end{equation}
where\cite{kallin}
\begin{equation}
G_{n_1 n_2} (\hat{q}) = \left( \frac{n_1!}{n_2 !} \right)^{1/2} 
              \left( \frac{-i q}{\sqrt{2} } \right)^{n_2 - n_1}
           L_{n_1}^{n_2-n_1} \left( \frac{ |q|^2}{2} \right) \mbox{\hspace{5mm},}
\end{equation}
with $\hat{q}=q_x + iq_y$, $q =|\hat{q}|$
 and $L^{m}_{n}$ being a generalized Laguerre polynomial. The only functions 
 relevant to our problem are 
\begin{eqnarray*}
& & G_{0,0} (\hat{q}) = \exp \left( -\frac{q^2}{4} \right) 
\mbox{\hspace{3mm},\hspace{3mm}}   
G_{0,1} (\hat{q}) = \frac{\hat{q}}{\sqrt{2}} \exp \left( -\frac{q^2}{4} \right)  
\mbox{\hspace{3mm},\hspace{3mm}}  \\
& & G_{1,1} (\hat{q}) = \left( 1 - \frac{q^2}{2} \right)\exp \left( -\frac{q^2}{4} \right) 
\end{eqnarray*}
 The modification 
of the Coulomb interaction due to the finite extent of the electron 
wave functions in $z$-direction is given by the expression:
\[
F_{\mu_1 \mu_4 \mu_2 \mu_3} (q) = 
 \int _{0}^{\infty} \int _{0}^{\infty} 
 dz_1 dz_2 e^{-q|z_1-z_2|}\chi _{\mu_1} (z_1) \chi _{\mu_4} (z_1)
                                    \chi _{\mu_2} (z_2) \chi _{\mu_3} (z_2)
 \mbox{\hspace{3mm}.}
\]
 We use the Fang-Howard approximation\cite{ando} with the parameter $b$ 
for the envelope functions of the  first and second subbands: 
\begin{eqnarray*}
& & \chi _1(z) = \sqrt{\frac{b^3}{2}} z \exp\left( -\frac{1}{2} bz \right) 
                                                           \mbox{\hspace{3mm},} \\
& & \chi _2(z) = \sqrt{\frac{b^5}{6}} z\left( z- \frac{3}{b} \right) 
                          \exp\left( -\frac{1}{2} bz \right)
                                                            \mbox{\hspace{3mm},}
\end{eqnarray*}
where the function $\chi _2(z)$ is found from the condition that 
  $\chi _1(z)$  and $\chi _2(z)$  are orthogonal. 

The first problem we are interested in is the ground state of the system. 
It is clear that if the splitting $\Delta $ between the layers is larger 
then the 
 characteristic inter-electron Coulomb interaction ($\Delta >1$) then the 
 ground state is a completely occupied level with $\mu =1$. Similarly, 
 if $~\Delta < -1$, the ground state is a completely occupied level 
 with $~\mu =2$. With changing $\Delta $ the transition
 between these two phases occurs. This transition can be either 
 sharp or smooth
 and in the last case the transition occur through the new intermediate phase. 
 The example of sharp phase transition in the two-level electron system with 
 the total filling factor $\nu =1$ is given in Ref.~\onlinecite{quinn}, where 
 the two 
  levels correspond to different Landau levels  of the same subband but 
 with opposite spin. There is also an example of a smooth 
 phase transition in the two-level system of composite fermions with filling
 factor $\nu =1$.\cite{apalkov} The two levels in Ref.~\onlinecite{apalkov}
  are the same 
 as the levels in Ref.~\onlinecite{quinn}. The only difference from 
 Ref.~\onlinecite{quinn} is in the type of interaction between the 
 particles. Our system is different from 
 the systems considered in Refs.~\onlinecite{quinn,apalkov} in two aspects: 
 the levels in 
 our case correspond to different subbands and the spin for both levels 
 is the same, i.e., we have  an additional exchange interaction 
 between the electrons in the different levels. 

 To find the ground state of the system in the intermediate range 
 of $\Delta $
 we follow the method of Ref.~\onlinecite{quinn} and use the standard 
 Hartree-Fock approximation. We assume the non-zero average values of 
 $\left< C^{+} _{k, 1 } C _{k, 1 }\right> $, 
$\left< C^{+} _{k, 2 } C _{k, 2 }\right> $ and
 $\left< C^{+} _{k, 1 } C _{k+Q, 2 }\right> $ which do not depend on $k$.  
 In general, the parameter $Q$ which  minimizes the Hartree-Fock Hamiltonian
 is not equal to zero.  
  For each value of $k$ we introduce the new 
 wavefunctions which are the eigenfunctions  of the Hartree-Fock Hamiltonian.  
 The creation and annihilation operators corresponding to these functions 
are $a^{+} _{k,i }$, 
 $a  _{k,i }$, $i=1,2$. The new functions are related to the original 
 ones by the rotation by angle $\theta $: 
\begin{equation}
 \begin{array}{l}
C_{k,1} =  a _{k,1} \cos(\theta /2) + a _{k,2} \sin(\theta /2) 
 \mbox{\hspace{3mm},} \\
C_{k+Q,2} =- a _{k,1} \sin(\theta /2) + a _{k,2} \cos(\theta /2)
 \mbox{\hspace{3mm}.} 
\end{array}
\end{equation}
 Assuming that $a_{k,1}$ corresponds to the lowest energy state we write the 
 average: $\left<  a^{+}_{k,i}a_{k,j} \right> = \delta _{i1}\delta _{j1}$.
Substituting expression (5) into the Hamiltonian (1) we get the Hartree-Fock
 Hamiltonian as a function of $\theta $ and $Q$: 
\begin{eqnarray*}
H^{HF} & = & \frac{1}{2} \left[- \Delta  
              - \epsilon ^{11}_{11} \cos^2 \left(\theta /2 \right)
              - \epsilon ^{01}_{12} \sin^2 \left(\theta /2 \right)
                               \right] \cos^2 \left(\theta /2 \right)  \\
     &  & + \frac{1}{2} \left[   \Delta 
              - \epsilon ^{00}_{22} \sin^2 \left(\theta /2 \right)
              - \epsilon ^{01}_{12} \cos^2 \left(\theta /2 \right)
                               \right] \sin^2 \left(\theta /2 \right)   \\
     &  & + \sin^2 \left(\theta /2 \right) \cos^2 \left(\theta /2 \right)
            V_{12} (Q)           \mbox{\hspace{3mm},}
\end{eqnarray*}
where $\epsilon ^{i_1 i_2}_{j_1 j_2}$ is the exchange energy of an electron 
 in the $i_1$th Landau level of the $j_1$th subband interacting with the 
 electrons of the same spin in the filled $i_2$th Landau level of the 
 $j_2$th subband:
\begin{equation}
\epsilon ^{i_1 i_2}_{j_1 j_2}  =  
     \int _0 ^{\infty }dk  F_{j_1 j_2 j_1 j_2}(k) \left| 
G _{i_1 i_2}(\hat{k}) \right|^2   
      \exp\left ( -\frac{k^2}{2} \right)   \mbox{\hspace{3mm}.}
\end{equation}
The potential $V_{12} (Q)$ is given by the expression:
\[
V_{12}(Q) = 
    \int _0 ^{\infty }dk \left[ F_{1 1 2 2}(k)  G _{1 1}(\hat{k}) 
G _{0 0}(\hat{k})  J_{0}(kQ) 
                       -  F_{1 2 1 2}(k)  \left|G _{1 2}(\hat{k}) 
\right|^2 J_{2}(kQ) \right]
      \exp\left ( -\frac{k^2}{2} \right) \mbox{\hspace{3mm}.}
\]

Analyzing the Hartree-Fock Hamiltonian, one can show that the energy 
minimum can be achieved only at $\theta =0$ or $\theta =\pi$, which 
corresponds to a completely occupied level $\mu =1$ or $\mu =2$, 
respectively. This means that
  there are  only two different phases with the a sharp transition 
 between them. Comparing the Hartree-Fock energy at $\theta =0$ 
 and $\theta =\pi$, we obtain the 
 inter-level splitting $\Delta ^{*} $ at which the transition between two 
 phases occurs:
\begin{equation}
\Delta ^{*} = \frac{1}{2} \left( \epsilon ^{00}_{22} - 
                                 \epsilon ^{11}_{11} \right)
              \mbox{\hspace{3mm}.}
\end{equation}
For $\Delta > \Delta ^{*}$ the ground state is the completely occupied 
 level with $\mu =1$ and for $\Delta < \Delta ^{*}$ the ground state is 
 the completely occupied  level with $\mu =2$. 

 To find the energy spectrum of neutral excitations we follow the method proposed 
 in Ref.~\onlinecite{macdonald} for a double layer system at electron filling 
 factor equal to 1. We introduce the normalized exciton operator, which 
creates 
 an electron in the empty level and a hole in the filled level: 
\begin{equation}
{\cal A}^{+} (\vec{q}) = \frac{1}{\sqrt{N}}  \sum _{k} e^{i q_x k} 
C^{+} _{k+q_y/2, 2} 
    C _{k-q_y/2, 1}   \mbox{\hspace{3mm},} 
\end{equation}
 where  $\vec{q}=(q_x,q_y)$ is the momentum of the electron-hole pair, $N$ 
is the 
 number of electrons  and we assume that the level with 
$\mu =1$ is filled. Then we calculate the 
energy spectrum of such excitations as an average of the Hamiltonian (1) in 
the state ${\cal A}^{+}(\vec{q} )\left| 0 \right>$:
\begin{eqnarray}
E(q) & = &  \left< {\cal A} (\vec{q}) H {\cal A}^{+}(\vec{q} ) \right> = 
     \Delta + \epsilon ^{11}_{11} - \epsilon ^{01}_{12} + 
           \frac{q}{2} F_{1212}(q) \exp \left( -q^2/2 \right)  \nonumber \\
   & &   - \int _{0}^{\infty } dk \left( 1- \frac{q^2}{2} \right) J_{0} (kq) 
                                     F_{1122}(k) \exp \left( -k^2/2 \right) 
 \nonumber \\
      & = &     (\Delta -\Delta ^{*})  + 
               \frac{1}{2} \left( \epsilon ^{00}_{22} + \epsilon ^{11}_{11} 
\right) 
             - \epsilon ^{01}_{12} + 
           \frac{q}{2} F_{1212}(q) \exp \left( -q^2/2 \right)    \nonumber \\
  & &    - \int _{0}^{\infty } dk \left( 1- \frac{q^2}{2} \right) J_{0} (kq) 
                                     F_{1122}(k) \exp \left( -k^2/2 \right) 
                                           \mbox{\hspace{3mm},} 
\end{eqnarray}
where we rewrite the $E(q)$ in terms of the difference $(\Delta -\Delta ^{*})$.
Because the Hamiltonian (1) does not conserve the number of excitons 
 we should take into account the processes which involve 
 several excitons.  Such processes are important at small 
 values of the level splitting $\Delta $. To take into account the 
 multi-exciton states we approximate  the Hamiltonian (1)  by\cite{macdonald}
\begin{equation}
H = \sum _{\vec{q}} \left\{   E(q) {\cal A}^{+}  (\vec{q} )  
{\cal A} (\vec{q} )  
              + \frac{1}{2} \left[   \lambda(\vec{q}) 
{\cal A} (\vec{q}){\cal A}(-\vec{q}) 
              +  \lambda ^{*}(\vec{q}) {\cal A}^{+} (\vec{q}) {
\cal A}^{+}(-\vec{q})
                               \right]
                    \right\}  \mbox{\hspace{3mm}.}
\end{equation}
The function $\lambda (q)$ is defined by the equation:
\begin{eqnarray}
\lambda (\vec{q}) & = & \left< H {\cal A}^{+} (\vec{q}) 
 {\cal A}^{+}(-\vec{q} ) \right> = 
           e^{2i\phi }  \frac{q}{2} F_{1212}(q) \exp \left( -q^2/2 \right) 
 \nonumber  \\
   & & -  e^{2i\phi }  \int _{0}^{\infty } dk \frac{q^2}{2} J_{2} (kq) 
                                     F_{1212}(k) \exp \left( -k^2/2 \right)  
                                \mbox{\hspace{3mm},}      
\end{eqnarray}
where $\phi = \arctan (q_y/q_x)$. After diagonalizing the Hamiltonian (10) 
 using  
 the Bogoliubov transformation\cite{kittel} we obtain the energy spectrum 
 of the 
elementary neutral excitations: 
\begin{equation}
E_{0}(q) = \sqrt{ E^2 (q) - |\lambda (\vec{q})|^2 }  \mbox{\hspace{3mm}.}
\end{equation}

If the level $\mu =2$ is occupied in the ground state, i.e. $\Delta
  < \Delta ^{*}$, 
then  the  Eqs.~(11) and (12) are the same and the only difference is 
in Eq.~(8) where 
we should write $-(\Delta -\Delta ^{*})=
|\Delta -\Delta ^{*}|$  instead of $(\Delta -\Delta ^{*})$.

The numerical calculations show that the lowest energy excitation (Eq. (12)) 
has a zero momentum. The energy of this excitation  
can be considered as a renormalized splitting, $\Delta _r$, between the 
 levels of 
 the double-level electron system.  It follows from Eqs.~(9), (11) and 
(12) that
\begin{equation}
\Delta _r = E_{0} (0) = |\Delta - \Delta ^{*} | +
               \frac{1}{2} \left( \epsilon ^{00}_{22} + \epsilon ^{11}_{11} \right) 
             - \epsilon ^{01}_{12} 
    - \int _{0}^{\infty } dk \left( 1- \frac{q^2}{2} \right)
                                     F_{1122}(k) \exp \left( -k^2/2 \right)
                       \mbox{\hspace{3mm}.}
\end{equation}
The splitting $\Delta _r$ reaches its lowest value at $\Delta = \Delta ^{*}$. 
The renormalization of the excitation energy in a 2D electron system  
 ($g$-factor enhancement) due to 
the inter-electron interaction was well studied both theoretically and
experimentally.\cite{g1,g2,g3,g4}
At the same time there are excitations of electron systems which are not 
altered 
 by the electron-electron
interaction. For example, Kohn theorem \cite{kohn} tells us that the energy of
a cyclotron excitation does not depend on the electron density. 

 In Fig.2(a) the critical level splitting $\Delta ^{*}$ is shown as a 
 function of 
 the Fang-Howard parameter $b$ by a solid line. With increasing $b$ the 
 spreading
 of the electron wave function in $z$ direction becomes smaller. This 
 results in increasing $\Delta ^{*}$, which can be understood from the 
 Eqs.~(6) and (7).
 The first term in equation (6), $\epsilon ^{00}_{22}$, is determined  by 
 inter-electron interaction in the first Landau level of the second 
 subband, while the second term $\epsilon ^{11}_{11}$ is determined by 
  the inter-electron interaction in the second Landau level of the first
  subband. 
 Since the spreading of the electrons in the second Landau level is broader 
 than the 
 spreading of the electrons  in the first Landau level, the 2D form-factor 
 of the inter-electron interaction, which is given by $G_{i_1,i_2}(q)$ 
 (Eq.~(4)),
 is larger for the electrons in the first Landau level.  Similarly, due to 
 a different spreading of the wave functions in $z$-direction for the electrons
 in the different subbands the form-factor $F_{\mu_1 \mu_2 \mu_3 \mu_4}(q)$ is
 larger for the electron in the first subband.  The competition 
 between these two tendencies gives rise to the increase of $\Delta ^*$ with 
 increasing $b$. We can see from Fig.2a that at $b \approx 1.6$ the effects 
 of the different spreading in $z$-direction and in the $(x,y)$-plane cancel 
 each other, so that $\Delta ^{*} =0$.   

  The dotted line in Fig.2(a) shows the dependence of the renormalized 
 splitting 
 $\Delta _r$ on the parameter $b$ for $\Delta =\Delta ^{*}$. We can see 
 again that with decreasing of the spreading of electron wave function in $z$ 
 direction the splitting $\Delta _r$ increases. 
 This increase of $\Delta _r$ has the same reason as the increase 
 of $\Delta ^{*}$.
 
 In Fig.2(b) the dispersion curve $E_0(q)$ (Eq. (12)) is shown for 
 $\Delta =\Delta ^{*}$ and 
 for different values of parameter $b$:  $b=1$ (solid line) and $b=2$ 
  (dotted line). We can see again that the excitation energy 
 increases with increasing  $b$. The effect becomes 
 stronger for larger values of momentum $q$. 
  
  In Fig.2(c) the dispersion curve $E_0(q)$ is shown for $b=1$ and for two different
 values of $\Delta $: for $\Delta = \Delta ^{*}$ (solid line) and for 
   $\Delta = \Delta ^{*} +0.05$ (dotted line). The increase of $\Delta $ has 
 no significant effect on the form of the dispersion curve due to a 
 small value of the function $\lambda (\vec{q})$. As a result the dispersion curve 
$E_0(k)$  has an almost linear dependence on splitting $\Delta $ for 
all values of momentum $q$ and can be approximated by $E_0 (q) \approx E(q)$.

\section{Phonon Spectroscopy}

  In this section we study the interaction of the quasi-2D 
 double subband electron system, discussed in the previous section, 
 with acoustic phonons. We consider the case of the 
 absorption phonon spectroscopy when the 2D electrons 
 are subjected to an external beam of non-equilibrium acoustic 
 phonons.\cite{phonon1,phonon2,phonon3,phonon4}

 Below we are using the isotropic Debye approximation in which 
 the phonon frequency has a linear dependence on the wave vector:
\[
\omega _{j} (K) = s_{j} K  \mbox{\hspace{3mm},}
\]
where $s_{j}$ is the speed of sound, $j$ is labeling the phonon modes,
 $j=1$ for longitudinal mode and $j=2,3$ for two transverse modes.
The wave vector $K$ is the three dimensional (3D) vector, 
 $K=|\vec{K}|$. For convenience we label the 3D
 vectors by the capital letters and their projections 
  by the corresponding small letters, $\vec{K}=(\vec{k},k_z)$.

  The electron-phonon interaction Hamiltonian can be written in terms
 of exciton operators ${\cal A}^+,{\cal A} $ (Eq. (8)):
\begin{equation}
H_{e-ph} = - \sum _{j, \vec{Q}} \frac{M_j (\vec{Q})}{\sqrt{V}} Z(q_z) \left[ 
          {\cal A}(\vec{q}) \hat{d}^{+}_j (\vec{Q}) + 
          {\cal A}^{+}(\vec{q}) \hat{d} _j (\vec{Q}) \right]
\mbox{\hspace{3mm},}
\end{equation}
where $\hat{d}^{+}_j$ is the creation operator of a phonon in the
$j$th mode, $V$ is a normalization volume; 
 $M_j (\vec{Q})$ are the matrix elements of electron-phonon
 interaction which are determined by the deformation potential and 
 piezoelectric coupling. In GaAs the matrix elements $M_j (\vec{Q})$ 
 are usually written in the form \cite{benedict}:
\begin{equation}
M_j (\vec{Q}) = \sqrt{\frac {\hbar }{2 \rho_0 s Q}} 
  \left[ -\beta 
           \frac{Q_x Q_y \xi _{j,z} +Q_y Q_z \xi _{j,x} +
                           Q_z Q_x \xi _{j,y}}{Q^2}
    - i \Xi _0 (\vec{\xi}_{j} \cdot \vec{Q} ) \right]
\mbox{\hspace{3mm},}
\end{equation}
 where $\rho _0 $ is the mass density of GaAs, 
 $\beta $ and $\Xi _0$ are the parameters of piezoelectric
 and deformation potential couplings,\cite{levinson} respectively; 
 $\vec{\xi}_j$ is the polarization vector of the $j$th phonon mode. 

 The form-factor $Z(q_z)$ in Eq. (14) is determined by the electron
 spreading in $z$ direction: 
\begin{equation}
Z(q_z) = \int dz e^{i q_z z} \chi_1(z) \chi_2(z)  \mbox{\hspace{3mm}.}
\end{equation}

To illustrate the specific feature of the electron-phonon interaction
 in our two-subband system we study the total rate of absorption of 
 phonons from an external pulse by the 2D electrons. At low 
enough temperature the rate of absorption is given by\cite{benedict}
\begin{equation}
\omega _{abs} = \frac{2\pi}{\hbar}  \sum _{j}
\int \frac{d \vec{Q}}{(2\pi )^3} \delta (E_0(q) - s Q) n_{j}(\vec{Q})
              \left| M_j (\vec{Q}) Z(q_z) \right|^2 R_{01}(q)
\mbox{\hspace{3mm},}
\end{equation}
where $ n_{j}(\vec{Q})$ is the phonon distribution function. 
The function $R_{01}(q)$ corresponding to the
transition from the first to the second Landau level is given by
\begin{equation}
R_{01}(q) = \left| G_{0,1}(\hat{q}) \right|^2 = 
 \frac{q^2}{2} e^{-q^2/2}  \mbox{\hspace{3mm}.}
\end{equation}

In Eq.~(17) we assume that the energy relaxation time of the 2D 
electron system is much shorter 
 than the characteristic time of the phonon absorption, which as shown 
 below is of  the order  of $10^{-10}~$s.    

From the expression for $R_{01}(q)$ (Eq.~(18)) one can see that the 
phonon absorption occurs mostly for the 2D  momentum $q$ being in  
the range of $q\sim 1\div 2$. In this range of $q$ the energy spectrum 
$E_0(q)$ is almost dispersionless (Fig.2(b)),
$E_0(q)\sim 0.08 + |\Delta -\Delta ^*|$ for $b=1$. Energy 
 conservation requires the energy of absorbed phonon to be
 equal to $E_0(q)$, i.e. $s\sqrt{q^2+q_z^2}=E_0(q)$. 
 Taking into account that the phonon momentum $q_z$ is restricted 
 by the  parameter $b$ ($|q_z|<b$) and that the speed of sound is about
 0.03 for longitudinal and 0.02 for transverse phonons, we see that 
 the  phonon absorption is suppressed in the two-subband system.
 This suppression is completely due to the  many-body effects, since
 the many-body gap, $E_0(q)\epsilon _C$, is greater than  $\hbar s /l$. 

 Let us assume that the non-equilibrium 
 phonon distribution function is isotropic. Then, substituting 
 Eqs.~(15) and (18) into the Eq.~(17) and performing the integration, we get:
\begin{eqnarray}
\frac{1}{\tau _{abs}} & = &   \frac{\omega _{abs}}{n(E_0(q_0))}  \nonumber \\ 
      & =  &    \frac{q_1^5}{s_1^2}  \int_0^1 du (1-u^2) 
                \left| Z(uq_1) \right|^2 
           \left[ M_d + 
                 \frac{9}{8} \frac{M_p}{q_1^2} u^2 (1-u^2)
           \right] \exp (-q_1^2(1-u^2)/2)     \nonumber \\
  & & +      \frac{q_2^3}{s_2}  \frac{M_p}{8} 
                \int_0^1 du (1-u^2)^2 (9 u^4 -2 u^2 +1) 
                 \left| Z(uq_2) \right|^2 
                  exp(-q_2^2(1-u^2)/2)
\mbox{\hspace{3mm},} 
\end{eqnarray} 
where $q_1 =E_0(q_0)/s_1$, $q_2 =E_0(q_0)/s_2$ and $q_0$ is taken on the 
 plateau of the dispersion curve $E_0(q)$ (Fig.2(b)).  
Here the constants $M_d$ and $M_p$, characterizing the deformation and 
 piezoelectric couplings, are:
\[
M_d = \frac{1}{4 \pi} \left( \frac{\Xi }{\varepsilon _C} \right)^2
    \frac{1 }{l^5 \rho _0}   \mbox{\hspace{3mm},}
\]
\[
M_p = \frac{1}{4 \pi} \left( \frac{\beta l }{\varepsilon _C} \right)^2
    \frac{ 1 }{l^5 \rho _0}   \mbox{\hspace{3mm}.}
\]

In Fig.3 the absorption rate is shown as a function of $\Delta -\Delta^{*}$ for
 the two values of parameter $b$: $b=1$ (solid line) and $b=1.5$ (dotted line). 
 Both curves have a single-peak structure and show 
 a weak dependence on $b$.  The energy $E_0(q_0)$ is 
 increasing with increasing $b$ (Fig.2(b)).  This results in the decrease of the 
 phonon absorption rate. However, the increase of $b$ is favorable for  
 the absorption of phonons with larger $q_z$, because the form-factor
 $Z(q_z)$ effectively cuts off the phonon momentum $q_z$ at 
 $q_z =b$. These two factors counteract each other,  
 which results in a weak absorption rate dependence on parameter $b$.
 For the focused phonon beam, the first effect becomes 
 more important and the decrease of the absorption rate with increasing $b$
 should be observed.
 
 The rate of the phonon absorption  for the non-interaction electron system 
is shown in Fig.3 by the dashed line. The main difference between the 
 interacting and non-interacting systems is in the shape of the dependencies 
 of the absorption rate on the level splitting. These dependencies have 
 the double-peak and single-peak structures for non-interacting and  
 interacting systems, respectively. For the non-interaction system the 
 excitation spectrum is just the bare gap $|\Delta |$ and the excitation 
 energy has any value starting from 0. The spectrum becomes gapless
 at $\Delta =0$. In this case  the maximum of 
 the absorption rate corresponds to the condition: $|\Delta | \sim s/l$.
 This results in a double-peak structure for the non-interaction 
 system.\cite{misha1} For the interacting system there is an energy gap 
  for any value of $\Delta $. Thus, the absorption rate as a function of 
 $\Delta $ has a single-peak structure.

 If the electron filling factor $\nu $ is changed from $\nu =0$ to $\nu =1$ 
 one should expect the transformation of the double-peak structure 
 into a single-peak one, because at small filling factor the inter-electron 
 interaction becomes less important. 
 This transformation occurs as the merging of the two peaks into 
 a single peak. Assuming that the interaction gap of the 2D electron system 
 is proportional to $\nu $, one can see that with increasing $\nu $ 
  the separation between peaks decreases linearly, and the two peaks 
 transform into a single peak at $\nu \approx 0.75$ for $b =1$. 
 The strength of the peaks increases linearly with increasing $\nu $, 
 because the phonon absorption 
 rate is proportional to the  electron density.  The transformation of 
 the double-peak into a single-peak structure with increasing electron
 filling factor can be observed experimentally. 
 The splitting between the
 levels $\Delta $ is the difference between the intersubband 
 splitting and cyclotron energy with some corrections due to
 exchange interaction with the background electrons. 
 Therefore, this splitting can be changed by changing 
 magnetic field. 

 At the same time the gate voltage should be adjusted to keep the electron 
 filling factor constant.  However, even for the changing filling
 factor at fixed electron density one should expect the single-peak 
 structure of the phonon absorption curve as a function of magnetic field 
if the electron filling factor is close to $\nu =1$.  The double-peak 
 structure appears when the electron filling factor is smaller then 
$\approx 0.7$. For a constant electron density the single-peak and 
 double-peak structures are both asymmetric when the difference
 $\Delta - \Delta ^*$ changes sign.

\section{Summary}
 We have studied a two-subband electron system in a quantizing 
 magnetic field at the total filling
 factor $\nu _0 =3$. When the second Landau level of the first 
 subband is close in energy to the first Landau level of the 
 second subband the system becomes effectively a two-level electron 
 system with filling factor $\nu =1$. Depending on  the value of 
 splitting between the two levels the system can be in one of the two 
 possible ground states:
 for $\Delta > \Delta ^*$ all the electrons  are in the first subband 
 and for $\Delta < \Delta ^*$ all the electrons are in the second subband. 
 The inter-electron interaction renormalizes the critical 
 splitting $\Delta ^*$ which should be zero for a non-interacting system. 
  Another effect related to the electron-electron interaction is  
 a non-zero gap for any value of level splitting $\Delta $. This effect 
 should be important for the phonon spectroscopy experiments.
 We show that due to the non-zero many-body gap the phonon absorption 
 rate has a single-peak structure as a function of the splitting 
 between the levels or as a function of magnetic field.
 At small electron filling factor $\nu $, when 
 the system can be considered as a non-interacting system, the double-peak 
 structure is expected.   
 The transformation of the double-peak structure into a single-peak one 
 occurs when the filling factor is changed from $\nu =0$ to $\nu =1$. 
 Our calculations show that the two peaks  merge at filling 
 factor $\nu \approx 0.7$.  

\section{Acknowledgments}

 This work was supported by the UK EPSRC.

\begin{figure}
\begin{center}
\begin{picture}(110,70)
\put(0,0){\includegraphics{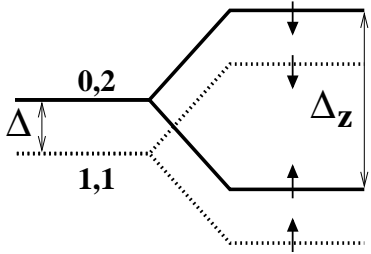}}
\end{picture}
\vspace*{2.0cm}
\caption{Single-electron energy levels of the two-subband system.
 Each level is numerated by the pair of numbers:
  the Landau level number and the subband number.
 The arrows indicate spin directions. 
 The interlevel splitting is $\Delta $. Zeeman splitting is $\Delta _z$.
}
%\label{fig1}
\end{center}
\end{figure}

\begin{figure}
\begin{center}
\begin{picture}(110,70)
\put(0,0){\includegraphics{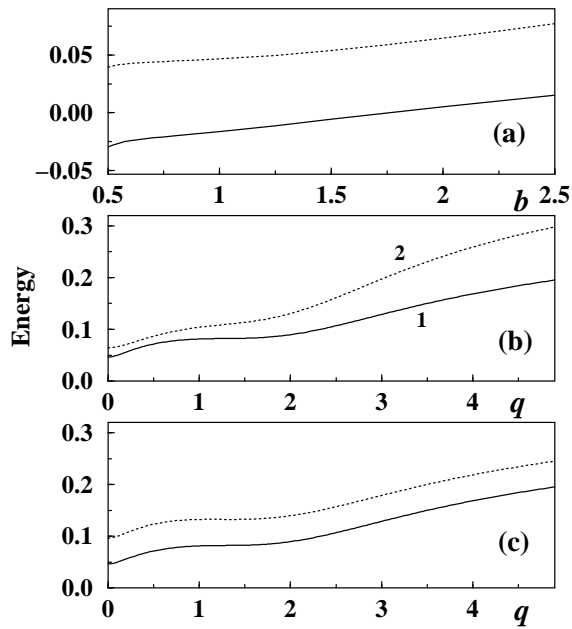}}
\end{picture}
\vspace*{7.0cm}
\caption{(a) The critical level splitting $\Delta ^*$ (Eq. (7)) and the 
 renormalized splitting $\Delta _r$ at $\Delta = \Delta ^*$ (Eq. (13)) are 
 shown as the functions of the Fang-Howard parameter $b$ by solid and dotted lines,
 respectively. (b) The dispersion curve $E_0 (q)$ (Eq. (12)) is shown 
 for $\Delta =\Delta ^*$. The numbers near the lines are the values of
 the parameter  $b$. (c) The dispersion 
 curve $E_0 (q)$ (Eq. (11)) is shown for $b=1$ and 
 different  $\Delta $: $\Delta =\Delta ^*$ (solid line) 
 and $\Delta =\Delta ^* + 0.05$ (dotted line). Both $q$ and $b$ are
 in units of inverse magnetic length, $1/l$; the energy 
 is in units of $\varepsilon _C$. 
}
%\label{fig2}
\end{center}
\end{figure}

\begin{figure}
\begin{center}
\begin{picture}(110,70)
\put(0,0){\includegraphics{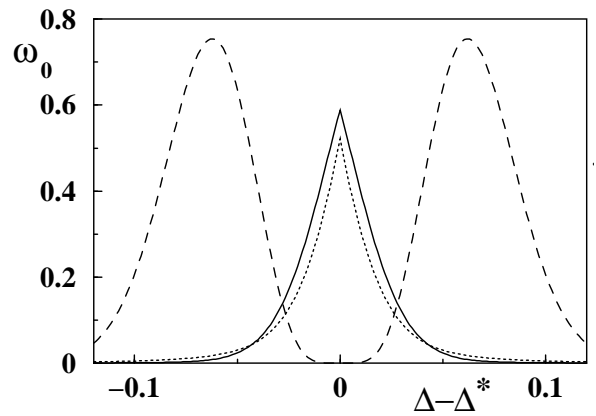}}
\end{picture}
\vspace*{3.0cm}
\caption{The phonon absorption rate $\omega _0= 1/\tau _{abs}$ (Eq. (19))
 as a function of $\Delta - \Delta ^*$ for $b=1$ (solid 
 line) and $b=1.5$ (dotted line). The absorption rate
 for the non-interacting system is shown by the dashed line. 
 The absorption rate $ \omega _0 $ is in units of $10^{10}~$s$^{-1}$, 
 $\Delta - \Delta ^*$ is in units of $\varepsilon _C$.
}
%\label{fig3}
\end{center}
\end{figure}

\end{document}